\documentclass[journal]{IEEEtran}
\ifCLASSINFOpdf
\else
\fi
\usepackage{amssymb}
\usepackage{hyperref}
\usepackage{amsmath,bm}
\usepackage{algorithm}
\usepackage{algpseudocode}
\usepackage{cite}
\usepackage{subfig}
\usepackage{tabularx}
\usepackage{tabulary}
\usepackage{mathrsfs}
\usepackage{graphicx}
\usepackage{epstopdf}
\usepackage{multirow}
\usepackage{makecell}
\usepackage{booktabs}
\usepackage{CJK}

\newtheorem{proposition}{Proposition}
\begin{document}
\title{Partial Maximum Correntropy Regression for Robust Electrocorticography Decoding}
\author{Yuanhao~Li,
	Badong~Chen,~\IEEEmembership{Senior~Member,~IEEE,}
	Gang~Wang,~\IEEEmembership{Member,~IEEE,}
	\\
	Natsue~Yoshimura,
	and~Yasuharu~Koike
	\thanks{This work was supported in part by the Japan Society for the Promotion of Science (JSPS) KAKENHI under Grant 19H05728, in part by the Japan Science and Technology Agency (JST) PRESTO (Precursory Research for Embryonic Science and Technology) under Grant JPMJPR17JA, in part by the Japan Science and Technology Agency (JST) SPRING (Support for Pioneering Research Initiated by Next Generation) under Grant JPMJSP2106, and in part by the National Natural Science Foundation of China under Grant U21A20485 and Grant 61976175. \emph{(Corresponding author: Yuanhao Li.)}}
	\thanks{Yuanhao Li, Natsue Yoshimura, and Yasuharu Koike are with the Institute of Innovative Research, Tokyo Institute of Technology, Yokohama 226-8503, Japan (correspondence e-mail: li.y.ay@m.titech.ac.jp).}
	\thanks{Badong Chen is with the Institute of Artificial Intelligence and Robotics, Xi'an Jiaotong University, Xi'an 710049, China.}
	\thanks{Gang Wang is with the Key Laboratory of Biomedical Information Engineering of Ministry of Education, School of Life Science and Technology, Institute of Biomedical Engineering, Xi'an Jiaotong University, Xi'an 710049, China.}} 
\maketitle

\begin{abstract}
	The \emph{Partial Least Square Regression} (PLSR) exhibits admirable competence for predicting continuous variables from inter-correlated brain recordings in the brain-computer interface. However, PLSR is in essence formulated based on the least square criterion, thus, being non-robust with respect to noises. The aim of this study is to propose a new robust implementation for PLSR. To this end, the maximum correntropy criterion (MCC) is used to propose a new robust variant of PLSR, called as \emph{Partial Maximum Correntropy Regression} (PMCR). The half-quadratic optimization is utilized to calculate the robust projectors for the dimensionality reduction, and the regression coefficients are optimized by a fixed-point approach. We evaluate the proposed PMCR with a synthetic example and the public Neurotycho electrocorticography (ECoG) datasets. The extensive experimental results demonstrate that, the proposed PMCR can achieve better prediction performance than the conventional PLSR and existing variants with three different performance indicators in high-dimensional and noisy regression tasks. PMCR can suppress the performance degradation caused by the adverse noise, ameliorating the decoding robustness of the brain-computer interface.
\end{abstract}


\section{Introduction}
\label{sec:introduction}
\IEEEPARstart{B}{rain}-computer interface (BCI) has been conceived as a promising technology that translates cerebral recordings generated by cortical neurons into appropriate commands for controlling neuroprosthetic devices \cite{wolpaw2002brain}. The capability of BCI for repairing or reproducing sensory-motor functions has been increasingly intensified by recent scientific and technological advances \cite{donoghue2002connecting,mussa2003brain,lebedev2006brain}. The noninvasive recordings, especially electroencephalogram (EEG) and magnetoencephalogram (MEG), are widely exploited to structure BCI systems due to their ease of use and satisfactory temporal resolution, whereas the noninvasive BCI systems could be limited in their capabilities and customarily require considerable training \cite{amiri2013review}. Invasive single-unit activities and local field potentials usually provide better decoding performance, which suffer pessimistic long-term stability, however, due to capriciousness in the recorded neuronal-ensembles \cite{chestek2007single}. A sophisticated alternative that exhibits higher signal amplitudes than EEG while presents superior long-term stability compared to invasive modalities, is the semi-invasive electrocorticography (ECoG) \cite{buzsaki2012origin}. Many studies in recent years have investigated the potentials of ECoG recording in decoding motions \cite{levine2000direct,leuthardt2004brain,ball2009differential,pistohl2008prediction,chin2007identification,chao2010long,shimoda2012decoding}. The serviceability of ECoG signal for online practice was also demonstrated in \cite{leuthardt2004brain,leuthardt2006electrocorticography,schalk2008two}.

The \emph{partial least square regression} (PLSR), which has been widely utilized in ECoG decoding tasks, is in particular suited for inter-correlated and high-dimensional problems, where the number of variables is larger than that of the observations. For instance, in \cite{chao2010long}, the authors successfully predicted the three-dimensional continuous hand trajectories of two monkeys during asynchronous food-reaching tasks from the time–frequency representations of subdural ECoG signals by PLSR algorithm. They further demonstrated the admirable prediction capability of PLSR in an epidural ECoG study \cite{shimoda2012decoding}. Although PLSR was originally developed for econometrics and chemometrics \cite{wold1966estimation}, it has emerged as a prevailing approach for neuroimaging and decoding \cite{krishnan2011partial}.

PLSR algorithm answers the regression problems primarily with dimensionality reduction techniques both on independent and dependent variables, in which the dimensionality-reduced samples (usually called as \emph{latent variables}) of respective sets exhibit maximal correlation, thus structuring association from independent to dependent variables. Multi-way PLSR, NPLSR in short, was proposed as a generalization for tensor variables \cite{bro1996multiway}. Recently, further studies have been investigated so as to ameliorate the performance of PLSR, most of which proposed to calculate the projectors or the regression coefficients with an additive regularization item \cite{chun2010sparse,eliseyev2012l1,eliseyev2016penalized,foodeh2020regularized}. Nevertheless, PLSR and the regularized variants essentially employ the least square criterion which assigns superfluous importance to the deviated noises, as a result, leading to poor robustness.

In the present study, we aim to propose a new robust version of PLSR by introducing the \emph{maximum correntropy criterion} (MCC) to replace the non-robust least square criterion, which was proposed in the \emph{information theoretic learning} (ITL) \cite{principe2010information}, and has achieved state-of-the-art robust approaches in different tasks, including regression \cite{liu2007correntropy,chen2012maximum,feng2015learning}, classification \cite{singh2014c,ren2018correntropy}, principal component analysis \cite{he2011robust}, and feature extraction \cite{dong2017correntropy}. 

Recently, a rudimentary implementation of the MCC in the PLSR algorithm was investigated in \cite{mou2018maximum}, in which the authors proposed to employ the MCC in the process of dimensionality reduction. However, the proposed algorithm in this study could be limited in some respects. First, except for the MCC-based dimensionality reduction, it remains computing the regression coefficients under the least square criterion. Second, it only executes the dimensionality reduction on the explanatory matrix. Thus, one has to estimate the regression coefficients separately for each dimension of the output matrix, which means it could be inadequate for multivariate response.

By comparison, in the present study, we desire to realize a more comprehensive implementation of the MCC framework in PLSR. The main contributions of this study are summarized as follows.

\begin{itemize}
\item[1)]
We reformulate PLSR thoroughly with the MCC framework, that not only the dimensionality reduction, but also the regression relations between the different variables are established by the MCC framework.
\item[2)]
Both the explanatory matrix and the response matrix are processed with dimensionality reduction. As a result, the proposed algorithm is adequate for multivariate response prediction.
\item[3)]
We use Gaussian kernel functions with individual kernel bandwidths for different reconstruction errors and prediction errors. Furthermore, each kernel bandwidth could be calculated from the corresponding set of errors.
\end{itemize}

The remainder of this paper is organized as follows. Section \ref{sec:plsr} introduces the mathematical derivation of the conventional PLSR algorithm. In Section \ref{sec:method}, we present a brief introduction about MCC, and reformulate PLSR with MCC. In Section \ref{sec:exp}, we evaluate the proposed method on synthetic and real ECoG datasets, respectively. We discuss about the proposed method in Section \ref{sec:disc}. Finally, the conclusion of this paper is given in Section \ref{sec:con}.

\section{Partial Least Square Regression}
\label{sec:plsr}

Consider a data set with the explanatory matrix $\mathbf{X}\in \mathbb{R}^{L\times N}$ and the response matrix $\mathbf{Y}\in \mathbb{R}^{L\times M}$, in which $N$ and $M$ denote the respective numbers of dimension, and $L$ is the number of observations. PLSR is an iterative regression algorithm which executes dimensionality reduction on explanatory and response matrices, so that the resultant latent variables in each iteration exhibit maximal covariance. 

In the first iteration, the original matrices are employed as the current residual matrices, i.e. $\mathbf{X}_1=\mathbf{X}$ and $\mathbf{Y}_1=\mathbf{Y}$. PLSR calculates two projectors $\mathbf{w}_1\in \mathbb{R}^N$ and $\mathbf{c}_1\in \mathbb{R}^M$ to acquire the corresponding latent variables, denoted as $\mathbf{t}_1=\mathbf{X}_1\mathbf{w}_1$ and $\mathbf{u}_1=\mathbf{Y}_1\mathbf{c}_1$, by maximizing the covariance
\begin{equation}
\label{equ:plsr}
\underset{\lVert \mathbf{w}_1 \rVert _2=\lVert \mathbf{c}_1 \rVert _2=1}{\max}\mathbf{t}_1^T\mathbf{u}_1=\mathbf{w}_1^T\mathbf{X}_1^T\mathbf{Y}_1\mathbf{c}_1
\end{equation}
where $T$ means transpose and $\lVert \cdot \rVert _2$ denotes the $L_2$-norm. One solves the aforesaid problem by singular value decomposition (SVD) on $\mathbf{X}_1^T\mathbf{Y}_1$. Then, one computes the loading vector $\mathbf{p}_1$ by the least square criterion as
\begin{equation}
\label{equ:plsr_p}
\begin{split}
&\underset{\mathbf{p}_1}{\min} \lVert \mathbf{X}_1-\mathbf{t}_1\mathbf{p}_1^T \rVert _2^2\;\\
\Leftrightarrow\,\,&\mathbf{p}_1=\mathbf{X}_1^T\mathbf{t}_1/(\mathbf{t}_1^T\mathbf{t}_1) \;\\
\end{split}
\end{equation}
thus organizing the regression from $\mathbf{t}_1$ to $\mathbf{X}_1$. PLSR supposes linear association from $\mathbf{t}_1$ to $\mathbf{u}_1$ furthermore by calculating a regression scalar $b_1$ by the least square criterion as
\begin{equation}
\label{equ:plsr_b}
\begin{split}
&\underset{b_1}{\min} \lVert \mathbf{u}_1-\mathbf{t}_1b_1 \rVert _2^2  \;\\
\Leftrightarrow\,\,&b_1=\mathbf{u}_1^T\mathbf{t}_1/(\mathbf{t}_1^T\mathbf{t}_1)  \;\\
\end{split}
\end{equation}
The residual matrices for the next iteration are updated by
\begin{equation}
\label{equ:plsr_upd}
\begin{split}
&\mathbf{X}_{2}=\mathbf{X}_{1}-\mathbf{t}_1\mathbf{p}_1^T \; \\ 
&\mathbf{Y}_{2}=\mathbf{Y}_{1}-b_1\mathbf{t}_1\mathbf{c}_1^T \; \\ 
\end{split}
\end{equation}
Such procedures are repeated by PLSR for the optimal number of factors $S$, which is usually selected by cross-validation. One then collects the outcomes of each iteration, $\mathbf{T}=[\mathbf{t}_1,..,\mathbf{t}_S]\in \mathbb{R}^{L\times S}$, $\mathbf{P}=[\mathbf{p}_1,..,\mathbf{p}_S]\in \mathbb{R}^{N\times S}$, $\mathbf{B}=\text{diag}(b_1,..,b_S)\in \mathbb{R}^{S\times S}$, and $\mathbf{C}=[\mathbf{c}_1,..,\mathbf{c}_S]\in \mathbb{R}^{M\times S}$. As a result, one rewrites the decomposition of $\mathbf{X}$ and the predicted response $\mathbf{\hat{Y}}$ as
\begin{equation}
\label{equ:plsr_decomp}
\begin{split}
&\mathbf{X}=\mathbf{T}\mathbf{P}^T \; \\ 
&\mathbf{\hat{Y}}=\mathbf{T}\mathbf{B}\mathbf{C}^T \; \\ 
\end{split}
\end{equation}
Thus, the prediction relationship from $\mathbf{X}$ to $\mathbf{\hat{Y}}$ is structured as
\begin{equation}
\label{equ:plsr_pred}
\mathbf{\hat{Y}}=\mathbf{X}\mathbf{H}
\end{equation}
in which $\mathbf{H}=\mathbf{P}^{T+}\mathbf{B}\mathbf{C}^T \in \mathbb{R}^{N\times M}$, and $\mathbf{P}^{T+}$ is the pseudo-inverse of $\mathbf{P}^{T}$.

One notes that, the PLSR algorithm utilizes the least square criterion not only in (\ref{equ:plsr_p})(\ref{equ:plsr_b}) to build the regression relationship, but also in (\ref{equ:plsr}) for the calculation of the projectors. Maximizing the covariance could be rewritten as \cite{barker2003partial}
\begin{equation}
\label{equ:plsr_ls}
\underset{\lVert \mathbf{w} \rVert _2=\lVert \mathbf{c} \rVert _2=1}{\min}\sum_{l=1}^L{\left(\begin{array}{c}\lVert\mathbf{x}^l-\mathbf{x}^l\mathbf{w}\mathbf{w}^T\rVert^2\\
	+\lVert\mathbf{y}^l-\mathbf{y}^l\mathbf{c}\mathbf{c}^T\rVert^2\\
	+\lVert\mathbf{x}^l\mathbf{w}-\mathbf{y}^l\mathbf{c}\rVert^2\end{array}\right)}
\end{equation}
in which $\mathbf{x}^l$ and $\mathbf{y}^l$ denote the $l$-th observations for $\mathbf{X}$ and $\mathbf{Y}$, respectively. Note that the subscripts for the number of factors are omitted for the reason of simplicity. The connotation of (\ref{equ:plsr_ls}) could be interpreted as follows. The first and second items in the summation denote the quadratic reconstruction errors for the input and output, respectively. For example, $\mathbf{x}^l\mathbf{w}$ indicates the dimensionality-reduced input of the $l$-th observation, and thus $\mathbf{x}^l\mathbf{w}\mathbf{w}^T$ denotes the reconstruction. The third item means the prediction error between the latent variables. PLSR obtains two projectors by minimizing the sum of three quadratic errors, which is non-robust with respect to noises.

\section{Partial Maximum Correntropy Regression}
\label{sec:method}
\subsection{Maximum Correntropy Criterion}
\label{subsec:mcc}
The correntropy concept was developed in the field of ITL as a generalized correlation function of random processes \cite{santamaria2006generalized}, which measures the similarity and interaction between vectors in a kernel space. Correntropy associates with the information potential of quadratic Renyi’s entropy \cite{liu2007correntropy}, in which the data's probability density function (PDF) is estimated by the Parzen's window approach \cite{silverman1986density,parzen1962estimation}. The correntropy which evaluates the similarity between two arbitrary variables $A$ and $B$, which is denoted by $\mathcal{V}(A,B)$ in this paper, is 
\begin{equation}
\label{equ:correntropy}
\mathcal{V}(A,B)=E[k(A-B)]
\end{equation}
where $k(\cdot)$ is a kernel function satisfying the Mercer’s theory \cite{vapnik2013nature} and $E[\cdot]$ signifies the expectation operator. In the practical application, one calculates the correntropy with $L$ observations by the following empirical estimation 
\begin{equation}
\label{equ:correntropy_emp}
\hat{\mathcal{V}}(A,B)=\frac{1}{L}\sum_{l=1}^L{k(a_l-b_l)}
\end{equation}
where the Gaussian kernel function $g_\sigma(x)\triangleq\exp(-x^2/2\sigma^2)$ with kernel bandwidth $\sigma$ is widely used for the kernel function $k(\cdot)$, thus leading to
\begin{equation}
\begin{split}
\label{equ:correntropy_gau}
\hat{\mathcal{V}}(A,B)&=\frac{1}{L}\sum_{l=1}^L{g_\sigma(a_l-b_l)}\; \\ 
&=\frac{1}{L}\sum_{l=1}^L{\exp(-\frac{(a_l-b_l)^2}{2\sigma^2})}\; \\ 
\end{split}
\end{equation}

Maximizing the correntropy in (\ref{equ:correntropy_gau}), called as the \emph{maximum correntropy criterion} (MCC), exhibits numerous advantages. Correntropy is essentially a local similarity measure, that its value is chiefly determined along $A=B$, i.e. zero-value error. As a result, the effect of large error caused by adverse noise is alleviated, leading to better robustness. In addition, correntropy could extract sufficient information from observations, since it considers all the even moments of errors. Moreover, it closely relates to the $m$-estimation, which can be regarded as a robust formulation of Welsch $m$-estimator \cite{liu2007correntropy,huber2004robust}.

\subsection{Partial Maximum Correntropy Regression}
\label{subsec:pmcr}
Substituting the three least-square items in the conventional PLSR (\ref{equ:plsr_ls}) with the maximum correntropy yields 
\begin{equation}
\label{equ:pmcr_proj}
\underset{\lVert \mathbf{w} \rVert _2=\lVert \mathbf{c} \rVert _2=1}{\max}\sum_{l=1}^L{\left(\begin{array}{c}g_{\sigma_x}(\mathbf{x}^l-\mathbf{x}^l\mathbf{w}\mathbf{w}^T)\\
	+g_{\sigma_y}(\mathbf{y}^l-\mathbf{y}^l\mathbf{c}\mathbf{c}^T)\\
	+g_{\sigma_r}(\mathbf{x}^l\mathbf{w}-\mathbf{y}^l\mathbf{c})\end{array}\right)}
\end{equation}
where $\sigma_x$, $\sigma_y$, and $\sigma_r$ denote the Gaussian kernel bandwidths for $\mathbf{X}$-reconstruction errors, $\mathbf{Y}$-reconstruction errors, and the prediction errors, respectively.

Then, one transforms the vectors $(\mathbf{x}^l-\mathbf{x}^l\mathbf{w}\mathbf{w}^T)$ and $(\mathbf{y}^l-\mathbf{y}^l\mathbf{c}\mathbf{c}^T)$ into scalars, provided that the two projectors $\mathbf{w}$ and $\mathbf{c}$ are unit-length vectors, i.e. $\mathbf{w}^T \mathbf{w} =\mathbf{c}^T \mathbf{c} =1$
\begin{equation}
\label{equ:pmcrqrt}
\begin{split}
\sqrt{\lVert \mathbf{x}^l-\mathbf{x}^l\mathbf{w}\mathbf{w}^T \rVert ^2}&=\sqrt{\mathbf{x}^l\mathbf{x}^{lT}-\mathbf{x}^l\mathbf{w}\mathbf{w}^T\mathbf{x}^{lT}} \; \\ 
\sqrt{\lVert \mathbf{y}^l-\mathbf{y}^l\mathbf{c}\mathbf{c}^T \rVert ^2}&=\sqrt{\mathbf{y}^l\mathbf{y}^{lT}-\mathbf{y}^l\mathbf{c}\mathbf{c}^T\mathbf{y}^{lT}} \; \\ 
\end{split}
\end{equation}
Subsequently, one obtains the following optimization problem to acquire the projectors
\begin{equation}
\label{equ:pmcr_projqrt}
\underset{\lVert \mathbf{w} \rVert _2=\lVert \mathbf{c} \rVert _2=1}{\max}\sum_{l=1}^L{\left(\begin{array}{c}g_{\sigma_x}(\sqrt{\mathbf{x}^l\mathbf{x}^{lT}-\mathbf{x}^l\mathbf{w}\mathbf{w}^T\mathbf{x}^{lT}})\\
	+g_{\sigma_y}(\sqrt{\mathbf{y}^l\mathbf{y}^{lT}-\mathbf{y}^l\mathbf{c}\mathbf{c}^T\mathbf{y}^{lT}})\\
	+g_{\sigma_r}(\mathbf{x}^l\mathbf{w}-\mathbf{y}^l\mathbf{c})\end{array}\right)}
\end{equation}

After obtaining $\mathbf{w}$ and $\mathbf{c}$, one calculates the latent variables as the conventional PLSR by $\mathbf{t}=\mathbf{X}\mathbf{w}$ and $\mathbf{u}=\mathbf{Y}\mathbf{c}$. We then compute the loading vector $\mathbf{p}$ and the regression scalar $b$ with MCC by
\begin{equation}
\label{equ:pmcr_p}
\underset{\mathbf{p}}{\max}\sum_{l=1}^L{g_{\sigma_p}(\mathbf{x}^l-\mathbf{t}^l\mathbf{p}^T)}
\end{equation}
\begin{equation}
\label{equ:pmcr_b}
\underset{b}{\max}\sum_{l=1}^L{g_{\sigma_b}(\mathbf{u}^l-\mathbf{t}^lb)}
\end{equation}
in which $\mathbf{t}^l$ and $\mathbf{u}^l$ are the $l$-th elements for the latent variables $\mathbf{t}$ and $\mathbf{u}$, respectively. $\sigma_p$ and $\sigma_b$ are the corresponding kernel bandwidths. The residual matrices are then updated similarly.	

One will repeat such procedures for the optimal number of factors and collects the acquired vectors from each iteration to organize the matrices $\mathbf{T}$, $\mathbf{P}$, $\mathbf{B}$, and $\mathbf{C}$, as the original PLSR. Ultimately, the predicted response $\mathbf{\hat{Y}}$ can be obtained from $\mathbf{X}$ by the regression relationship (\ref{equ:plsr_pred}). The above-mentioned PLSR variant which is reformulated based on MCC, is called \emph{partial maximum correntropy regression} (PMCR).

In what follows, we discuss about the optimization, convergence analysis, and determination of hyper-parameters considering the proposed PMCR algorithm.

\subsubsection{Optimization}
\label{subsec:pmcr_opt}
Three optimization problems (\ref{equ:pmcr_projqrt})(\ref{equ:pmcr_p})(\ref{equ:pmcr_b}) need to be addressed in PMCR. Those two regression problems (\ref{equ:pmcr_p})(\ref{equ:pmcr_b}) can be well solved by a fixed-point method proposed in \cite{chen2015convergence}. We mainly consider the problem (\ref{equ:pmcr_projqrt}) for the projectors $\mathbf{w}$ and $\mathbf{c}$. Based on the \emph{half-quadratic} (HQ) optimization \cite{ren2018correntropy}, (\ref{equ:pmcr_projqrt}) could be rewritten as
\begin{equation}
\label{equ:pmcr_projqrt_hq}
\underset{\begin{array}{c}\lVert \mathbf{w} \rVert _2=\\ \lVert \mathbf{c} \rVert _2=1\end{array}}{\max}\sum_{l=1}^L{\left(\begin{array}{c}
	\sup\{\alpha_l\frac{\mathbf{x}^l\mathbf{x}^{lT}-\mathbf{x}^l\mathbf{w}\mathbf{w}^T\mathbf{x}^{lT}}{2\sigma_x^2}-\varphi(\alpha_l)\}\\
	+\sup\{\beta_l\frac{\mathbf{y}^l\mathbf{y}^{lT}-\mathbf{y}^l\mathbf{c}\mathbf{c}^T\mathbf{y}^{lT}}{2\sigma_y^2}-\varphi(\beta_l)\}\\
	+\sup\{\gamma_l\frac{(\mathbf{x}^l\mathbf{w}-\mathbf{y}^l\mathbf{c})^2}{2\sigma_r^2}-\varphi(\gamma_l)\}\end{array}\right)}
\end{equation}
by introducing three sets of auxiliaries $\{\alpha_l\}_{l=1}^L$, $\{\beta_l\}_{l=1}^L$, and $\{\gamma_l\}_{l=1}^L$, respectively, and $\varphi(\cdot)$ is a convex conjugated function of $g(\cdot)$. Hence we can conclude that optimizing (\ref{equ:pmcr_projqrt}) equals to updating $(\alpha_l,\beta_l,\gamma_l)$ and $(\mathbf{w},\mathbf{c})$ alternately by
\begin{equation}
\label{equ:pmcr_proj_j}
\underset{\begin{array}{c}\lVert \mathbf{w} \rVert _2=\\ \lVert \mathbf{c} \rVert _2=1,\\ \alpha_l,\beta_l,\gamma_l\in \mathbb{R}\end{array}}{\max}J\triangleq\sum_{l=1}^L{\left(\begin{array}{c}
	\alpha_l\frac{\mathbf{x}^l\mathbf{x}^{lT}-\mathbf{x}^l\mathbf{w}\mathbf{w}^T\mathbf{x}^{lT}}{2\sigma_x^2}-\varphi(\alpha_l)\\
	+\beta_l\frac{\mathbf{y}^l\mathbf{y}^{lT}-\mathbf{y}^l\mathbf{c}\mathbf{c}^T\mathbf{y}^{lT}}{2\sigma_y^2}-\varphi(\beta_l)\\
	+\gamma_l\frac{(\mathbf{x}^l\mathbf{w}-\mathbf{y}^l\mathbf{c})^2}{2\sigma_r^2}-\varphi(\gamma_l)\end{array}\right)}
\end{equation}

Since the HQ optimization is an iterative process, we denote the $k$-th HQ iteration with the subscript $k$. First, according to the HQ mechanism, we update the auxiliaries with the current projectors $(\mathbf{w}_{k},\mathbf{c}_{k})$ by
\begin{equation}
\label{equ:pmcr_proj_auxi}
\begin{split}
\alpha_{l,k+1}&=-\exp(-\frac{\mathbf{x}^l\mathbf{x}^{lT}-\mathbf{x}^l\mathbf{w}_{k}\mathbf{w}_{k}^T\mathbf{x}^{lT}}{2\sigma_x^2}) \; \\ 
\beta_{l,k+1}&=-\exp(-\frac{\mathbf{y}^l\mathbf{y}^{lT}-\mathbf{y}^l\mathbf{c}_{k}\mathbf{c}_{k}^T\mathbf{y}^{lT}}{2\sigma_y^2}) \; \\ 
\gamma_{l,k+1}&=-\exp(-\frac{(\mathbf{x}^l\mathbf{w}_{k}-\mathbf{y}^l\mathbf{c}_{k})^2}{2\sigma_r^2}) \; \\ 
(l&=1,..,L)\; \\ 
\end{split}
\end{equation}
Then, to optimize the projectors, we rewrite (\ref{equ:pmcr_proj_j}) by collecting the terms of projectors and omitting the auxiliaries as
\begin{equation}
\label{equ:pmcr_projqrt_hq_proj}
\begin{split}
\underset{\begin{array}{c}\lVert \mathbf{w} \rVert _2=\\ \lVert \mathbf{c} \rVert _2=1\end{array}}{\max}J_p\triangleq\sum_{l=1}^L{\left(\begin{array}{c}
	(\frac{\gamma_l}{2\sigma_r^2}-\frac{\alpha_l}{2\sigma_x^2})\mathbf{x}^l\mathbf{w}\mathbf{w}^T\mathbf{x}^{lT}\\
	+(\frac{\gamma_l}{2\sigma_r^2}-\frac{\beta_l}{2\sigma_y^2})\mathbf{y}^l\mathbf{c}\mathbf{c}^T\mathbf{y}^{lT}\\
	-\frac{\gamma_l}{\sigma_r^2}\mathbf{x}^l\mathbf{w}\mathbf{c}^T\mathbf{y}^{lT}\end{array}\right)}
\end{split}
\end{equation}
This exhibits a quadratic programming problem constrained by nonlinear equations, for which there exist numerous solutions in the literature, such as the sequential quadratic programming (SQP) that represents the state of the art in nonlinear programming methods \cite{fletcher2013practical}. The comprehensive procedures for PMCR are summarized in Algorithm \ref{algo_hqpmcr}.

\begin{algorithm}[h]
	\caption{Partial Maximum Correntropy Regression}
	\label{algo_hqpmcr}
	\begin{algorithmic}[1]
		\State \textbf{Input}:
		matrices of explanation $\mathbf{X}$ and response $\mathbf{Y}$;
		number of factors $S$; a small positive value $\varsigma$
		\State \textbf{Output}:
		prediction model $\mathbf{\hat{Y}}=\mathbf{X}\mathbf{H}$
		\State initialize $\mathbf{X}_1=\mathbf{X}$ and $\mathbf{Y}_1=\mathbf{Y}$;
		\For{$s=1,2,..,S$}
		\State initialize the projectors by the conventional PLSR;
		\State initialize $converged=$ FALSE; 
		\Repeat
		\State auxiliaries-step: update $(\alpha_l,\beta_l,\gamma_l)$ with (\ref{equ:pmcr_proj_auxi}); 
		\State projectors-step: update $(\mathbf{w}_s,\mathbf{c}_s)$ with (\ref{equ:pmcr_projqrt_hq_proj});
		\If{ the difference of the objective function (\ref{equ:pmcr_projqrt}) is smaller than $\varsigma$}
		\State $converged=$ TRUE
		\EndIf
		\Until $converged==$ TRUE
		\State compute latent variables $\mathbf{t}_s=\mathbf{X}_s\mathbf{w}_s$ and $\mathbf{u}_s=\mathbf{Y}_s\mathbf{c}_s$;
		\State compute $\mathbf{p}_s$ and $b_s$ (\ref{equ:pmcr_p})(\ref{equ:pmcr_b}) by the fixed-point method; 
		\State update the residual matrices $\mathbf{X}_{s+1}=\mathbf{X}_{s}-\mathbf{t}_s\mathbf{p}_s^T$ and $\mathbf{Y}_{s+1}=\mathbf{Y}_{s}-b_s\mathbf{t}_s\mathbf{c}_s^T$; 
		\EndFor
		\State organize the matrices $\mathbf{T}=[\mathbf{t}_1,..,\mathbf{t}_S]$, $\mathbf{P}=[\mathbf{p}_1,..,\mathbf{p}_S]$, $\mathbf{B}=\text{diag}(b_1,..,b_S)$, and $\mathbf{C}=[\mathbf{c}_1,..,\mathbf{c}_S]$; 
		\State compute $\mathbf{H}=\mathbf{P}^{T+}\mathbf{B}\mathbf{C}^T$
	\end{algorithmic}
\end{algorithm}

\subsubsection{Convergence Analysis}
\label{subsec:pmcr_con}
For the regression coefficients $\mathbf{p}$ and $b$, one could find the detailed convergence analysis in \cite{chen2015convergence}. Here we mainly consider the convergence of the projectors $\mathbf{w}$ and $\mathbf{c}$ in the optimization problem (\ref{equ:pmcr_projqrt}). Because correntropy is in nature an $m$-estimator \cite{liu2007correntropy}, the local optimums of (\ref{equ:pmcr_projqrt}) are close to the global optimum, which is proved in one theoretical study \cite{loh2015regularized}. Accordingly, in what follows, we analyze that (\ref{equ:pmcr_projqrt}) would converge to a local optimum with the HQ optimization.

\begin{proposition}
	\label{prop1}
	If we have $J_p(\mathbf{w}_{k},\mathbf{c}_{k})\leqslant J_p(\mathbf{w}_{k+1},\mathbf{c}_{k+1})$ fixing $(\alpha_l,\beta_l,\gamma_l)=(\alpha_{l,k+1},\beta_{l,k+1},\gamma_{l,k+1})$, the optimization problem with respect to the projectors in (\ref{equ:pmcr_projqrt}) would converge to a local optimum.
\end{proposition}

\emph{Proof:} The convergence is proved as
\begin{equation} 
\begin{split} \label{hq_con1}
&J(\mathbf{w}_{k},\mathbf{c}_{k},\alpha_{l,k},\beta_{l,k},\gamma_{l,k})\;\\
\leqslant &J(\mathbf{w}_{k},\mathbf{c}_{k},\alpha_{l,k+1},\beta_{l,k+1},\gamma_{l,k+1})\; \\
\leqslant &J(\mathbf{w}_{k+1},\mathbf{c}_{k+1},\alpha_{l,k+1},\beta_{l,k+1},\gamma_{l,k+1})\; \\
\end{split}
\end{equation}
where the first inequality is guaranteed by the HQ mechanism \cite{ren2018correntropy}, and the second inequality arises from the assumption of the present proposition. $\hfill\blacksquare$

Note that, to guarantee the convergence of the projectors in  (\ref{equ:pmcr_projqrt}), it is unnecessary to strictly attain the maximum of (\ref{equ:pmcr_projqrt_hq_proj}) at every projectors-step of Algorithm \ref{algo_hqpmcr}. On the contrary, so long as the updated projectors lead to a larger objective function $J_p$ at each projectors-step, the problem (\ref{equ:pmcr_projqrt}) can converge to local optimum. This reveals great convenience in practice, that one only needs a few SQP iterations. One can finish the projectors-step once confirming the increase on $J_p$, thus accelerating the convergence.

\subsubsection{Hyper-Parameter Determination}
\label{subsec:pmcr_hyp}
There exist five Gaussian kernel bandwidths $\sigma_x$, $\sigma_y$, $\sigma_r$, $\sigma_p$, and $\sigma_b$, respectively, to be determined in practice. In the literature, an effective method to acquire a suitable kernel bandwidth for probability density estimation, named as \emph{Silverman's rule}, was proposed in \cite{silverman1986density}. By denoting the current set of errors as $E$ with $L$ observations, the bandwidth is
\begin{equation}
\label{equ:silv}
\sigma^2=1.06\times\min\{\sigma_E,\frac{R}{1.34}\}\times(L)^{-1/5}
\end{equation}
in which $\sigma_E$ is the standard deviation of the $L$ errors, and $R$ denotes the interquartile range. 

\section{Experiments}
\label{sec:exp}
In this section, we evaluated the proposed PMCR algorithm on synthetic example and real ECoG datasets, respectively, by comparing it to existing PLSR methods. Specifically speaking, we compared PMCR to the conventional PLSR method. Next, we involved a regularized PLSR (RPLSR) for the comparison that was proposed in a recent work \cite{foodeh2020regularized}. Furthermore, PMCR was compared to the rudimentary MCC-based variant that was proposed in \cite{mou2018maximum}, which was denoted by MCC-PLSR in what follows. For a evenhanded comparison, all the algorithms used an identical number of factors, which would be selected by the conventional PLSR in five-fold cross-validation. The maximal number of factors was set as $100$.

Considering the performance indicators for the evaluation, we used three typical measures in regression tasks: i) Pearson's correlation coefficient (\emph{r}), ii) root mean squared error (RMSE) which is computed by
\begin{equation}
\label{equ:rmse}
\text{RMSE}=\sqrt{\frac{1}{L}\sum_{l=1}^{L}\lVert \mathbf{\hat{y}}_l-\mathbf{y}_l \rVert^2}
\end{equation}
in which $\mathbf{\hat{y}}_l$ and $\mathbf{y}_l$ are the $l$-th observations for the prediction $\mathbf{\hat{Y}}$ and the target $\mathbf{Y}$, respectively, and iii) mean absolute error (MAE) which represents the average $L_1$-norm distance
\begin{equation}
\label{equ:mae}
\text{MAE}=\frac{1}{L}\sum_{l=1}^{L}\lVert \mathbf{\hat{y}}_l-\mathbf{y}_l \rVert
\end{equation}

One notes that, to compare the robustness between different algorithms, only contaminating the training samples by noises with isolating testing data from contamination is an extensively approved and implemented method in the literature, as advised in \cite{zhu2004class}. Accordingly, only the training sample would suffer the adverse contamination in the following experiments.

\subsection{Synthetic Dataset}
\label{subsec:toy}
First, we considered inter-correlated, high-dimensional, and noisy synthetic example, in which various PLSR methods were assessed with different levels of contamination. Randomly we created 300 i.i.d. latent variables $\mathbf{t}\sim U(0,1)$ for training, and 300 i.i.d. latent variables $\mathbf{t}\sim U(0,1)$ for testing, in which $U$ denotes the uniform distribution, and the dimension of $\mathbf{t}$ was set as $20$. We generated the hypothesis from the latent variable to the explanatory and response matrices then. Specifically, we randomly generated the transformation matrices with arbitrary values, which were subject to the standard normal distribution. The latent variables $\mathbf{t}$ were multiplied with a $20\times 500$ transformation matrix, resulting in a $300\times 500$ explanatory matrix for input. Similarly, we used a $20\times 3$ transformation matrix to acquire a $300\times 3$ response matrix for output. Hence, we predicted the multivariate responses from $500$-dimensional explanatory variables with $300$ training samples, and evaluated the prediction performance on the other $300$ testing samples.              

\begin{figure*}[t!]
	\centering
	\includegraphics[width=0.78\textwidth]{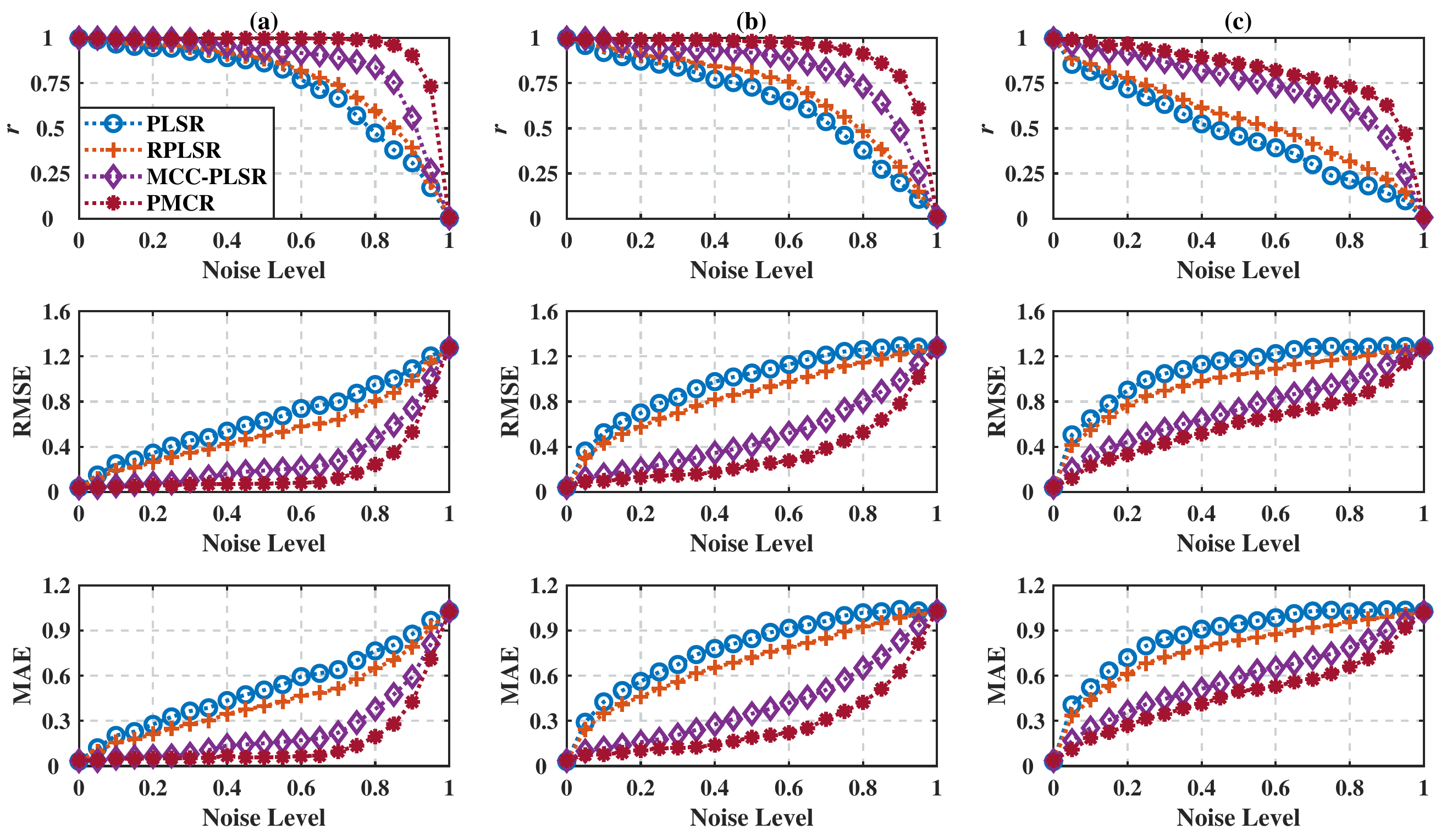}
	\caption{Average regression performance indicators of the inter-correlated, high-dimensional, and contaminated synthetic dataset under different noise standard deviations with noise levels from $0$ to $1.0$. (a): noise standard deviation $=30$, (b): noise standard deviation $=100$, and (c): noise standard deviation $=300$. The performance indicators were acquired from the $100$ Monte-Carlo repetitive trials and averaged across three dimensions of the output. The proposed PMCR algorithm realized better  performance than the existing PLSR algorithms consistently for \emph{r}, RMSE, and MAE, in particular when the training set was contaminated considerably.}
	\label{fig_toy}
\end{figure*}

Considering the contamination for the synthetic dataset, we supposed the explanatory matrix to be contaminated, since the adverse noise mainly happens to the brain recording, which is usually used as the explanatory in the BCI system. Therefore, some training samples were randomly selected, the inputs of which were then replaced by noises with large amplitude. The noise level denotes the proportion of the contaminated samples in the entirety, which was increased from $0$ to $1.0$ with a step $0.05$. For the distribution of noise, we utilized the zero-mean Gaussian distribution with large standard deviations to imitate outliers, where $30$, $100$, and $300$ were employed, respectively. We evaluated the algorithms with $100$ Monte-Carlo repetitive trials, and illustrate the results in Fig. \ref{fig_toy}. Note that, the results were further averaged across three dimensions of the output.
        
One observes from Fig. \ref{fig_toy} that for all the three different noise distributions, the proposed PMCR achieved superior prediction performance compared to the other existing PLSR algorithms consistently for \emph{r}, RMSE, and MAE, respectively, in particular when the training set suffered considerable contamination.

\subsection{ECoG Dataset}
\label{subsec:ecog}
To further demonstrate the superior robustness of the PMCR algorithm, we evaluated the various PLSR algorithms with the following practical brain decoding task. In this subsection, we used the public available Neurotycho ECoG dataset which was initially proposed in \cite{shimoda2012decoding}. 

\subsubsection{Dataset Description}
Two Japanese macaques, denoted by Monkey B and C, respectively, were commanded to capture the foods with their right hands, in which the continuous three-dimensional positions of hand with a sampling rate of 120 Hz were recorded by optical motion capture instrument. For both Monkey B and C, ten recording sessions were performed, and each recording session lasted 15 minutes. Both macaques were in advance implanted with customized 64-channel ECoG electrodes on the contralateral (left) hemisphere, which covered the regions from the prefrontal cortex to the parietal cortex. ECoG signals were recorded simultaneously during each session with a sampling rate of 1,000 Hz. In accordance with \cite{shimoda2012decoding}, for each recording session, the data of the first ten minutes was used to train a prediction model, while the data of the remaining five minutes was used to evaluate the prediction performance of the trained model. The schematic drawings of the experiments and ECoG electrodes are shown in Fig. \ref{fig_paradigm} (a) and (b), respectively.

\subsubsection{Decoding Paradigm}
\begin{figure*}[t!]
	\centering
	\includegraphics[width=0.78\textwidth]{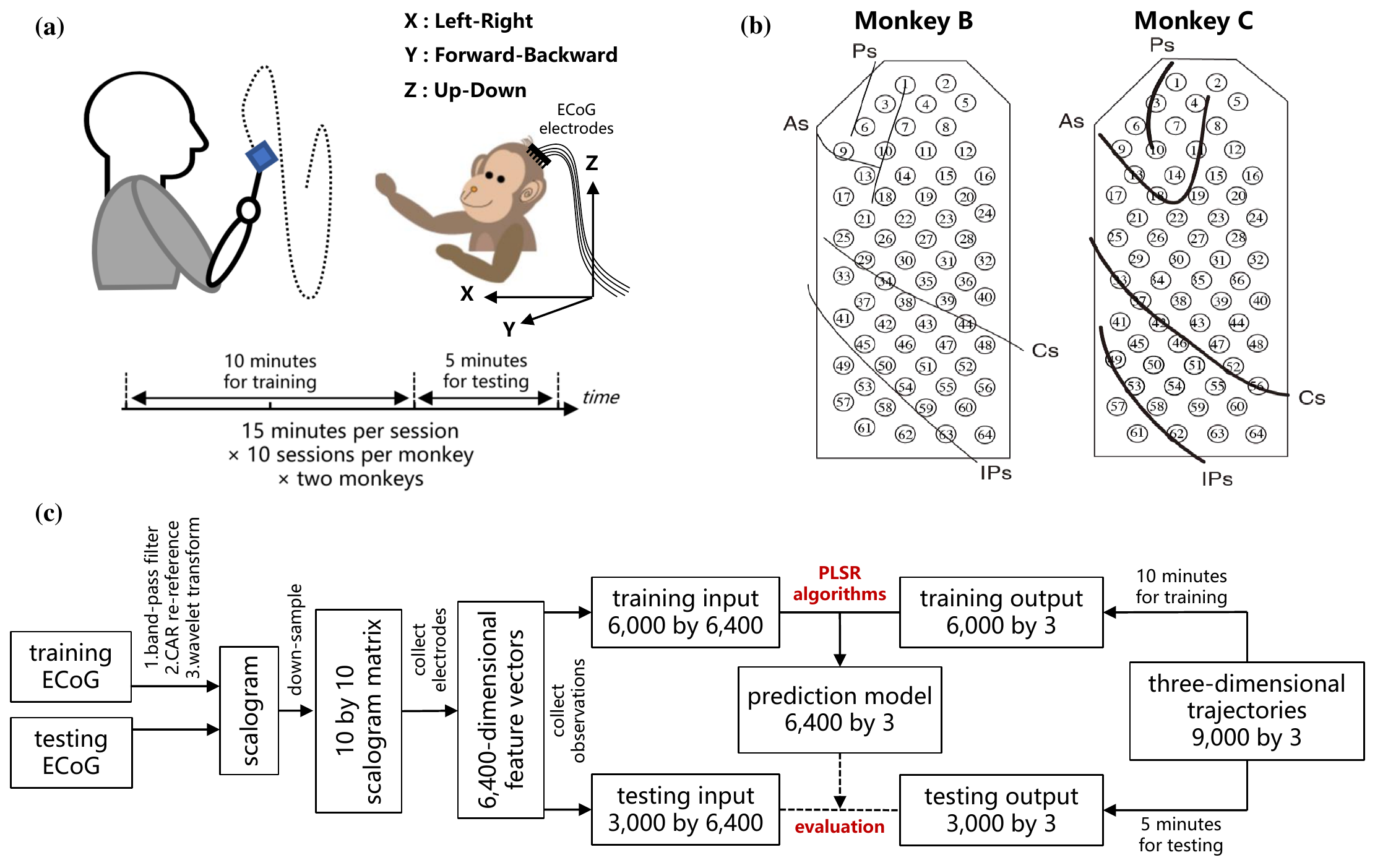}
	\caption{Experimental protocol of the Neurotycho ECoG dataset and decoding paradigm to evaluate the robustness of the different PLSR algorithms. (a): Macaques retrieved foods in a three-dimensional random location, in which the body-centered coordinates of the right wrists and ECoG signals were recorded simultaneously. (b): Both Monkey B and C were implanted with 64-channel epidural ECoG electrodes on the contralateral (left) hemisphere, overlaying the regions from prefrontal cortex to parietal cortex. Ps: principal sulcus, As: arcuate sulcus, Cs: central sulcus, IPs: intraparietal sulcus. (a) and (b) were modified from \cite{shimoda2012decoding}, which provides the details of this public dataset in \underline{http://neurotycho.org/epidural-ecog-food-tracking-task}. (c): Decoding diagram from ECoG signals to three-dimensional trajectories. The training ECoG signals are contaminated to assess the robustness of different algorithms.}
	\label{fig_paradigm}
\end{figure*}

Considering feature extraction, we used an identical offline decoding paradigm as in \cite{shimoda2012decoding}. Initially, ECoG signals were preprocessed by a tenth-order Butterworth bandpass filter with cutoff frequencies from 1 to 400 Hz, and then re-referenced by the common average referencing (CAR). The continuous three-dimensional trajectories of the right wrist were down-sampled to 10 Hz, thus, leading to 9,000 samples in one session (10 Hz $\times$ 60 sec $\times$ 15 min). The three-dimensional position of time \emph{t} was predicted from the ECoG signals in the previous one second. To describe the features of ECoG signals, we utilized the time–frequency representation method. For the position of time \emph{t}, the ECoG signals at each electrode from \emph{t} - 1.1 s to \emph{t} were processed by the Morlet wavelet transformation. Ten center frequencies ranging from 10 to 120 Hz with equal gaps on the logarithmic scale were considered for the wavelet transformation, overlaying the frequency bands which are most relevant to the motion tasks \cite{shimoda2012decoding}. The resultant time–frequency scalogram was then resampled at ten temporal lags with a 0.1 s gap (\emph{t} - 1 s, \emph{t} - 0.9 s,..., \emph{t} - 0.1 s). Therefore, the input variable of each sample exhibited a 6,400-dimensional vector (64 channels $\times$ 10 frequencies $\times$ 10 temporal lags), and the output variable was the three-dimensional position of the right hand. As a result, we trained a regression model with 6,000 samples (the first ten minutes) to predict the three-dimensional output variable from the 6,400-dimensional input variable, and evaluated the algorithms with other 3,000 testing samples (the remaining five minutes). The illustrative diagrams for ECoG decoding are summarized in Fig. \ref{fig_paradigm}  (c).

\begin{figure*}[t!]
	\centering
	\includegraphics[width=1.0\textwidth]{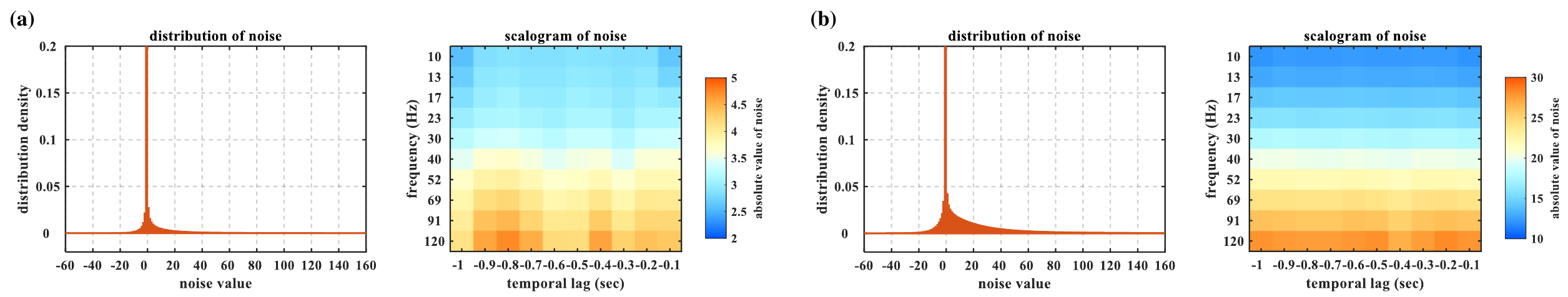}
	\caption{Distributions and scalograms of the resultant time–frequency noise resulting from the ECoG sampling noises. (a): noise level = $10^{-3}$ (the deteriorated proportion of training set = $0.6645\pm 0.0089$), (b): noise level = $10^{-2}$ (the deteriorated proportion of training set $\approx1$). The time–frequency noises are calculated by subtracting the training sets which are obtained from acoustic and contaminated ECoG signals, respectively. The distributions are averaged across 20 sessions of Monkey B and C, while the scalograms are averaged across electrodes. The peaks of distributions are truncated to emphasize the heavy-tailed characteristic.}
	\label{fig_ecog_noise}
\end{figure*}

\subsubsection{Contamination}
To demonstrate the robustness of different algorithms in the practical ECoG decoding task, the ECoG signals were artificially contaminated by outlier to simulate the detrimental artifact. To be specific, we stochastically selected a certain quantity of the training ECoG samplings and replaced them with outliers which were subject to a zero-mean Gaussian distribution with the variance $50$ times that of the signals from the corresponding channel. As stated in \cite{ball2009signal}, the blink related artifacts were remarkably found in ECoG signal, which exhibit much larger amplitudes than a normal ECoG recording. Hence, we used the above-mentioned approach to artificially generate adverse artifacts, so as to contaminate the ECoG signals. This method has been widely utilized in the literature to deteriorate the brain recordings for evaluating the robustness of different algorithms \cite{wang2011l1,chen2018common}.

Note that, for this ECoG dataset, the `Noise Level' signifies the ratio of the contaminated ECoG samplings in the entirety, which is different from the ratio of the deteriorated samples in 6,000 training samples. The proportion of the affected training samples can be evidently larger than the indicated noise level, since one contaminated ECoG sampling can deteriorate several time windows in the feature extraction. For example, when the noise level is indicated as $10^{-3}$, the deteriorated proportion of the training set is $(0.6645\pm 0.0089)$. Furthermore, we illustrate how the noises influence the time–frequency feature in Fig. \ref{fig_ecog_noise}. One could obviously perceive the heavy-tailed characteristic of the feature noise, which is particularly intractable for the least square criterion. In addition, the effects of high-frequency band are more prominent, due to the property of impulsive noise.

\subsubsection{Spatio-spectro-temporal Pattern}
Studying how spatio-spectro-temporal weights in the regression model contribute to the entirety can help investigate the neurophysiological pattern. The element of the trained prediction model $\mathbf{H}$ is denoted by $h_{\text{ch},\text{freq},\text{temp}}$, which corresponds to the ECoG electrode `ch',  the frequency `freq', and the temporal lag `temp'. Thus one could compute the spatio-spectro-temporal contributions by the ratio between the summation of absolute values of each domain and the summation of absolute values of the entire model 
\begin{equation}
\label{equ:cont_ch}
W_c(\text{ch})=\frac{\sum_{\text{freq}}{\sum_{\text{temp}}{|h_{\text{ch},\text{freq},\text{temp}}|}}}{\sum_{\text{ch}}{\sum_{\text{freq}}{\sum_{\text{temp}}{|h_{\text{ch},\text{freq},\text{temp}}|}}}}
\end{equation}
\begin{equation}
\label{equ:cont_freq}
W_f(\text{freq})=\frac{\sum_{\text{ch}}{\sum_{\text{temp}}{|h_{\text{ch},\text{freq},\text{temp}}|}}}{\sum_{\text{ch}}{\sum_{\text{freq}}{\sum_{\text{temp}}{|h_{\text{ch},\text{freq},\text{temp}}|}}}}
\end{equation}
\begin{equation}
\label{equ:cont_temp}
W_t(\text{temp})=\frac{\sum_{\text{ch}}{\sum_{\text{freq}}{|h_{\text{ch},\text{freq},\text{temp}}|}}}{\sum_{\text{ch}}{\sum_{\text{freq}}{\sum_{\text{temp}}{|h_{\text{ch},\text{freq},\text{temp}}|}}}}
\end{equation}
where $W_c(\text{ch})$, $W_f(\text{freq})$, and $W_t(\text{temp})$ denote the contributions of the ECoG electrode `ch',  the frequency `freq', and the temporal lag `temp', respectively.

\subsubsection{Results}
Firstly we assessed the different algorithms with the uncontaminated ECoG recordings. Hence, when the noise level was zero, the average performance indicators were gained from the acoustic 20 sessions (Monkey B and C). Afterwards, we contaminated each session with 5 repetitive trials. Thus, for each noise level, the different algorithms were assessed for 100 times  (20 sessions $\times$ 5 repetitive trials). In TABLE \ref{tab_eECoG}, we give the performance indicators for each algorithm under the noise levels $0$, $10^{-3}$, and $10^{-2}$, respectively. One could observe that, the proposed PMCR realized the optimal prediction capability consistently, except for the Y-axis under noise level $0$. Further, one can note that, when the noise level was zero, the proposed PMCR achieved better results than the other algorithms on the X-axis and Z-axis. The main reason is, in the acoustic sessions, motion-related artifacts have been considerably observed in the ECoG recordings \cite{shimoda2012decoding}, which further confirms the exceptional robustness of PMCR in practical brain decoding tasks.

\begingroup
\setlength{\tabcolsep}{5pt} 
\renewcommand{\arraystretch}{1.4}
\begin{table*}[t!]
	\centering
	\caption{Regression performance indicators of different algorithms on the Neurotycho epidural ECoG dataset with three noise levels $0$, $10^{-3}$, and $10^{-2}$, respectively. For the noise level $0$, the results were averaged across the acoustic 20 sessions (Monkey B and C). For the other noise levels, the average results were acquired from 100 trials (20 sessions $\times$ 5 repetitive trials). The results are given in mean$\pm$deviation, while the optimal results under each condition are marked in bold. The proposed PMCR realized the optimal prediction performance consistently, except for the Y-position with the noise level being $0$.}
	\label{tab_eECoG}
	\resizebox{0.85\textwidth}{!}{
		\begin{tabular}{p{2.0cm}<{\centering} p{2.0cm}<{\centering} p{2.00cm}<{\centering} p{2.68cm}<{\centering} p{2.68cm}<{\centering} p{2.68cm}<{\centering} p{2.68cm}<{\centering}}
			\toprule
			\hline
			\multicolumn{7}{c}{\multirow{1}{*}{\textbf{X-position}}}\\
			\hline
			\multicolumn{3}{c|}{\multirow{1}{*}{Algorithm}}& PLSR & RPLSR & MCC-PLSR &PMCR\\
			\hline
			\multicolumn{1}{c}{\multirow{9}{*}{Noise Level}}&\multicolumn{1}{|c|}{\multirow{3}{*}{$0$}}&\multicolumn{1}{c|}{\emph{r}}&0.4378$\pm$0.0933 & 0.4550$\pm$0.0925 & 0.4598$\pm$0.0942 &\textbf{0.4679$\pm$0.0947}\\
			\multicolumn{1}{c}{\multirow{9}{*}{}}&\multicolumn{1}{|c|}{\multirow{3}{*}{}}&\multicolumn{1}{c|}{RMSE}&0.9287$\pm$0.0810 & 0.9037$\pm$0.0653 & 0.8954$\pm$0.0809 &\textbf{0.8835$\pm$0.0786}\\
			\multicolumn{1}{c}{\multirow{9}{*}{}}&\multicolumn{1}{|c|}{\multirow{3}{*}{}}&\multicolumn{1}{c|}{MAE}&0.7026$\pm$0.0640 & 0.6872$\pm$0.0530 & 0.6749$\pm$0.0628 &\textbf{0.6658$\pm$0.0651}\\
			\cline{2-7}
			\multicolumn{1}{c}{\multirow{9}{*}{}}&\multicolumn{1}{|c|}{\multirow{3}{*}{$10^{-3}$}}&\multicolumn{1}{c|}{\emph{r}}&0.3334$\pm$0.1165 & 0.3558$\pm$0.1132 & 0.3684$\pm$0.1127 &\textbf{0.3873$\pm$0.1274}\\
			\multicolumn{1}{c}{\multirow{9}{*}{}}&\multicolumn{1}{|c|}{\multirow{3}{*}{}}&\multicolumn{1}{c|}{RMSE}&0.9729$\pm$0.0652 & 0.9543$\pm$0.0648 & 0.9397$\pm$0.0728 &\textbf{0.9276$\pm$0.0705}\\
			\multicolumn{1}{c}{\multirow{9}{*}{}}&\multicolumn{1}{|c|}{\multirow{3}{*}{}}&\multicolumn{1}{c|}{MAE}&0.7291$\pm$0.0756 & 0.7174$\pm$0.0689 & 0.7092$\pm$0.0786 &\textbf{0.6987$\pm$0.0759}\\
			\cline{2-7}
			\multicolumn{1}{c}{\multirow{9}{*}{}}&\multicolumn{1}{|c|}{\multirow{3}{*}{$10^{-2}$}}&\multicolumn{1}{c|}{\emph{r}}&0.1524$\pm$0.1399 & 0.1713$\pm$0.1353 & 0.1926$\pm$0.1342 &\textbf{0.2238$\pm$0.1382}\\
			\multicolumn{1}{c}{\multirow{9}{*}{}}&\multicolumn{1}{|c|}{\multirow{3}{*}{}}&\multicolumn{1}{c|}{RMSE}&1.0249$\pm$0.1105 & 1.0022$\pm$0.1097 & 0.9845$\pm$0.1129 &\textbf{0.9681$\pm$0.1094}\\
			\multicolumn{1}{c}{\multirow{9}{*}{}}&\multicolumn{1}{|c|}{\multirow{3}{*}{}}&\multicolumn{1}{c|}{MAE}&0.7655$\pm$0.1428 & 0.7485$\pm$0.1383 & 0.7396$\pm$0.1392 &\textbf{0.7246$\pm$0.1397}\\
			\hline
			\multicolumn{7}{c}{\multirow{1}{*}{\textbf{Y-position}}}\\
			\hline
			\multicolumn{3}{c|}{\multirow{1}{*}{Algorithm}}& PLSR & RPLSR & MCC-PLSR &PMCR\\
			\hline
			\multicolumn{1}{c}{\multirow{9}{*}{Noise Level}}&\multicolumn{1}{|c|}{\multirow{3}{*}{$0$}}&\multicolumn{1}{c|}{\emph{r}}&0.5426$\pm$0.1019 & \textbf{0.5582$\pm$0.1026} & 0.5547$\pm$0.1017 & 0.5549$\pm$0.1022\\
			\multicolumn{1}{c}{\multirow{9}{*}{}}&\multicolumn{1}{|c|}{\multirow{3}{*}{}}&\multicolumn{1}{c|}{RMSE}&0.8483$\pm$0.0969 & \textbf{0.8198$\pm$0.0951} & 0.8246$\pm$0.0948 & 0.8233$\pm$0.0952\\
			\multicolumn{1}{c}{\multirow{9}{*}{}}&\multicolumn{1}{|c|}{\multirow{3}{*}{}}&\multicolumn{1}{c|}{MAE}&0.6487$\pm$0.0762 & \textbf{0.6304$\pm$0.0796} & 0.6362$\pm$0.0744 & 0.6358$\pm$0.0759\\
			\cline{2-7}
			\multicolumn{1}{c}{\multirow{9}{*}{}}&\multicolumn{1}{|c|}{\multirow{3}{*}{$10^{-3}$}}&\multicolumn{1}{c|}{\emph{r}}&0.4114$\pm$0.1309 & 0.4284$\pm$0.1285 & 0.4425$\pm$0.1302 & \textbf{0.4602$\pm$0.1296}\\
			\multicolumn{1}{c}{\multirow{9}{*}{}}&\multicolumn{1}{|c|}{\multirow{3}{*}{}}&\multicolumn{1}{c|}{RMSE}&0.9188$\pm$0.0963 & 0.8962$\pm$0.0958 & 0.8795$\pm$0.0979 & \textbf{0.8608$\pm$0.1002}\\
			\multicolumn{1}{c}{\multirow{9}{*}{}}&\multicolumn{1}{|c|}{\multirow{3}{*}{}}&\multicolumn{1}{c|}{MAE}&0.6960$\pm$0.1007 & 0.6849$\pm$0.1014 & 0.6631$\pm$0.0983 & \textbf{0.6539$\pm$0.1021}\\
			\cline{2-7}
			\multicolumn{1}{c}{\multirow{9}{*}{}}&\multicolumn{1}{|c|}{\multirow{3}{*}{$10^{-2}$}}&\multicolumn{1}{c|}{\emph{r}}&0.2084$\pm$0.1514 & 0.2206$\pm$0.1489 & 0.2593$\pm$0.1502 & \textbf{0.2723$\pm$0.1537}\\
			\multicolumn{1}{c}{\multirow{9}{*}{}}&\multicolumn{1}{|c|}{\multirow{3}{*}{}}&\multicolumn{1}{c|}{RMSE}&0.9781$\pm$0.1143 & 0.9542$\pm$0.1117 & 0.9306$\pm$0.1159 & \textbf{0.9294$\pm$0.1146}\\
			\multicolumn{1}{c}{\multirow{9}{*}{}}&\multicolumn{1}{|c|}{\multirow{3}{*}{}}&\multicolumn{1}{c|}{MAE}&0.7354$\pm$0.1028 & 0.7173$\pm$0.1077 & 0.7086$\pm$0.1105 & \textbf{0.7043$\pm$0.1042}\\
			\hline
			\multicolumn{7}{c}{\multirow{1}{*}{\textbf{Z-position}}}\\
			\hline
			\multicolumn{3}{c|}{\multirow{1}{*}{Algorithm}}& PLSR & RPLSR & MCC-PLSR &PMCR\\
			\hline
			\multicolumn{1}{c}{\multirow{9}{*}{Noise Level}}&\multicolumn{1}{|c|}{\multirow{3}{*}{$0$}}&\multicolumn{1}{c|}{\emph{r}}&0.6320$\pm$0.0324 & 0.6395$\pm$0.0328 & 0.6482$\pm$0.0359 & \textbf{0.6504$\pm$0.0372}\\
			\multicolumn{1}{c}{\multirow{9}{*}{}}&\multicolumn{1}{|c|}{\multirow{3}{*}{}}&\multicolumn{1}{c|}{RMSE}&0.7968$\pm$0.0281 & 0.7814$\pm$0.0293 & 0.7747$\pm$0.0296 & \textbf{0.7628$\pm$0.0275}\\
			\multicolumn{1}{c}{\multirow{9}{*}{}}&\multicolumn{1}{|c|}{\multirow{3}{*}{}}&\multicolumn{1}{c|}{MAE}&0.6181$\pm$0.0222 & 0.6102$\pm$0.0280 & 0.6055$\pm$0.0241 & \textbf{0.5989$\pm$0.0265}\\
			\cline{2-7}
			\multicolumn{1}{c}{\multirow{9}{*}{}}&\multicolumn{1}{|c|}{\multirow{3}{*}{$10^{-3}$}}&\multicolumn{1}{c|}{\emph{r}}&0.4875$\pm$0.0708 & 0.4935$\pm$0.0701 & 0.5158$\pm$0.0857 & \textbf{0.5259$\pm$0.0814}\\
			\multicolumn{1}{c}{\multirow{9}{*}{}}&\multicolumn{1}{|c|}{\multirow{3}{*}{}}&\multicolumn{1}{c|}{RMSE}&0.9272$\pm$0.0712 & 0.9129$\pm$0.0682 & 0.8958$\pm$0.0742 & \textbf{0.8834$\pm$0.0738}\\
			\multicolumn{1}{c}{\multirow{9}{*}{}}&\multicolumn{1}{|c|}{\multirow{3}{*}{}}&\multicolumn{1}{c|}{MAE}&0.6932$\pm$0.0800 & 0.6894$\pm$0.0814 & 0.6804$\pm$0.0852 & \textbf{0.6645$\pm$0.0782}\\
			\cline{2-7}
			\multicolumn{1}{c}{\multirow{9}{*}{}}&\multicolumn{1}{|c|}{\multirow{3}{*}{$10^{-2}$}}&\multicolumn{1}{c|}{\emph{r}}&0.2399$\pm$0.1185 & 0.2456$\pm$0.1173 & 0.2615$\pm$0.1148 & \textbf{0.2803$\pm$0.1186}\\
			\multicolumn{1}{c}{\multirow{9}{*}{}}&\multicolumn{1}{|c|}{\multirow{3}{*}{}}&\multicolumn{1}{c|}{RMSE}&1.0168$\pm$0.0804 & 0.9917$\pm$0.0785 & 0.9605$\pm$0.0842 & \textbf{0.9485$\pm$0.0809}\\
			\multicolumn{1}{c}{\multirow{9}{*}{}}&\multicolumn{1}{|c|}{\multirow{3}{*}{}}&\multicolumn{1}{c|}{MAE}&0.7532$\pm$0.0883 & 0.7429$\pm$0.0892 & 0.7208$\pm$0.0893 & \textbf{0.7146$\pm$0.0887}\\
			\hline
			\bottomrule
	\end{tabular}}
\end{table*}

In addition, we studied how the neurophysiological patterns of different algorithms were influenced by the sampling noises. We examined the changes between the spatio-spectro-temporal weights which were attained from the acoustic sessions and the contaminated sessions, respectively. Fig. \ref{fig_cont} illustrates how the neurophysiological patterns were influenced by the noise level $10^{-3}$ for each algorithm. The prediction model of Monkey B's Z-position was employed here. In Fig. \ref{fig_cont} (a), for each algorithm we plot the spatial patterns that were attained from the acoustic sessions and the contaminated sessions, respectively, and also quantify how much the spatial pattern was affected by showing the summation of the absolute values of the difference between the weights of all electrodes, i.e. $\sum{|W_c(\text{ch})-W_c'(\text{ch})|}$, where $W_c(\text{ch})$ denotes the spatial weight of the acoustic sessions and $W_c'(\text{ch})$ is obtained from the contaminated sessions. In Fig. \ref{fig_cont} (b), for the spectral pattern, we show $W_f(\text{freq})$ and $W_f'(\text{freq})$ for each algorithm, which were obtained from the acoustic and contaminated sessions, respectively, and show the summation of absolute value of the difference $\sum{|W_f(\text{freq})-W_f'(\text{freq})|}$. For the temporal pattern, similarly we illustrate $W_t(\text{temp})$ and $W_t'(\text{temp})$ for each algorithm in Fig. \ref{fig_cont} (c), which were attained from the acoustic and contaminated sessions, respectively. We also give $\sum{|W_t(\text{temp})-W_t'(\text{temp})|}$ for each algorithm. One sees that, the proposed PMCR algorithm realized the minimum summation of the absolute value of the difference between the original pattern and the affected pattern for each domain. This demonstrates the robustness of PMCR to extract more accurate neurophysiological information.

\begin{figure*}[t!]
	\centering
	\includegraphics[width=1.0\textwidth]{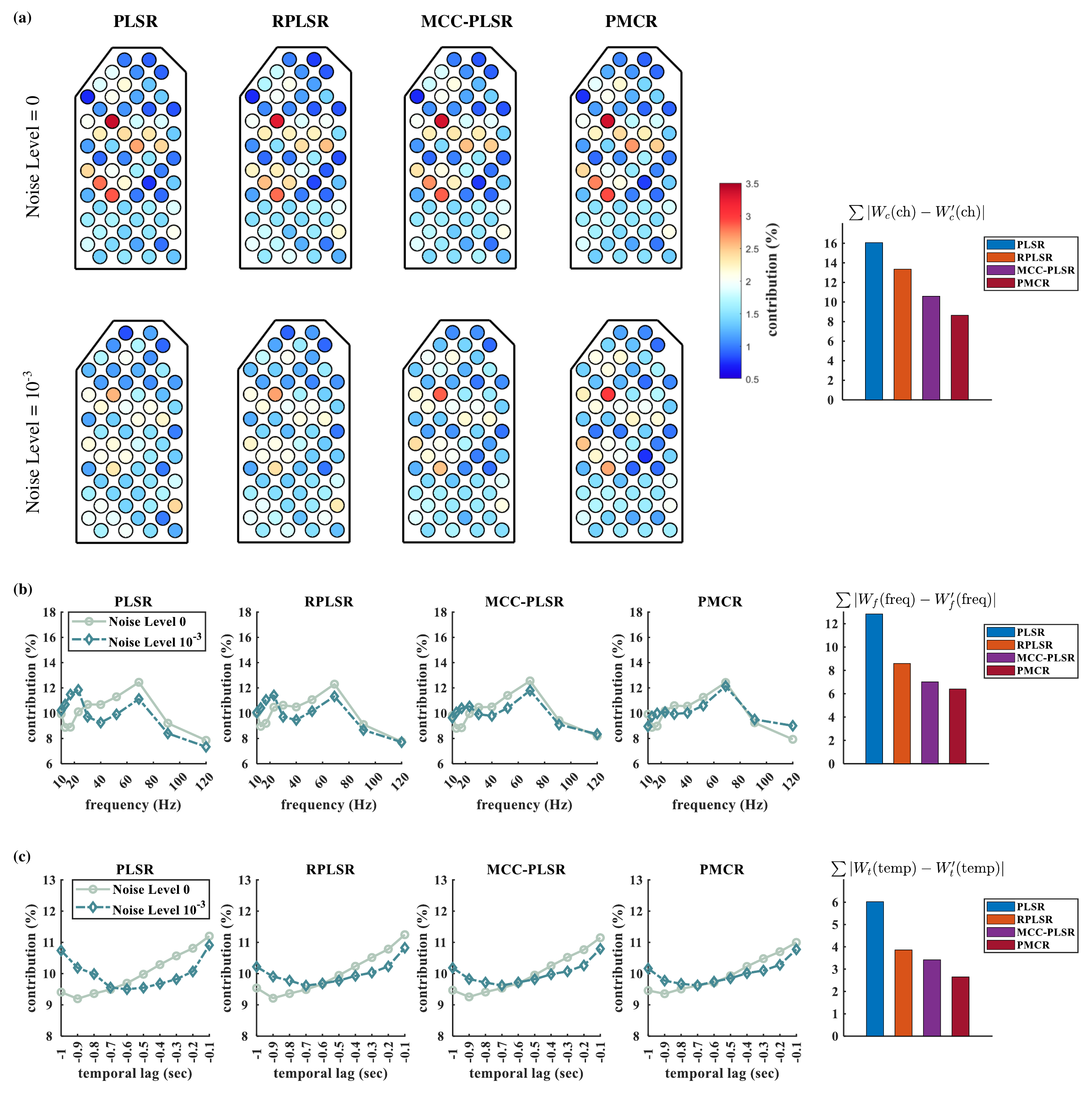}
	\caption{Spatio-spectro-temporal contributions of the prediction model for Monkey B's Z-position under noise levels $0$ and $10^{-3}$. (a): spatial contributions for each electrode. The upper row shows the spatial contributions of the acoustic sessions $W_c(\text{ch})$ and the bottom row presents the contributions of the contaminated sessions $W_c'(\text{ch})$. (b): spectral contributions for each frequency. (c): temporal contributions for each temporal lag. For both (b) and (c), the contributions of the acoustic sessions are shown in solid lines, while the contributions of the contaminated sessions are illustrated in dashed lines. For each domain, the summations of the absolute value of the difference between the original pattern and the deteriorated pattern are given for each algorithm on the right. PMCR achieved the minimum difference between the original patterns and the deteriorated patterns for each domain.}
	\label{fig_cont}
\end{figure*}

\section{Discussion}
\label{sec:disc}
\subsection{Number of Factors}
\label{subsec:n_fac}

\begin{figure*}[t!]
	\centering
	\includegraphics[width=0.78\textwidth]{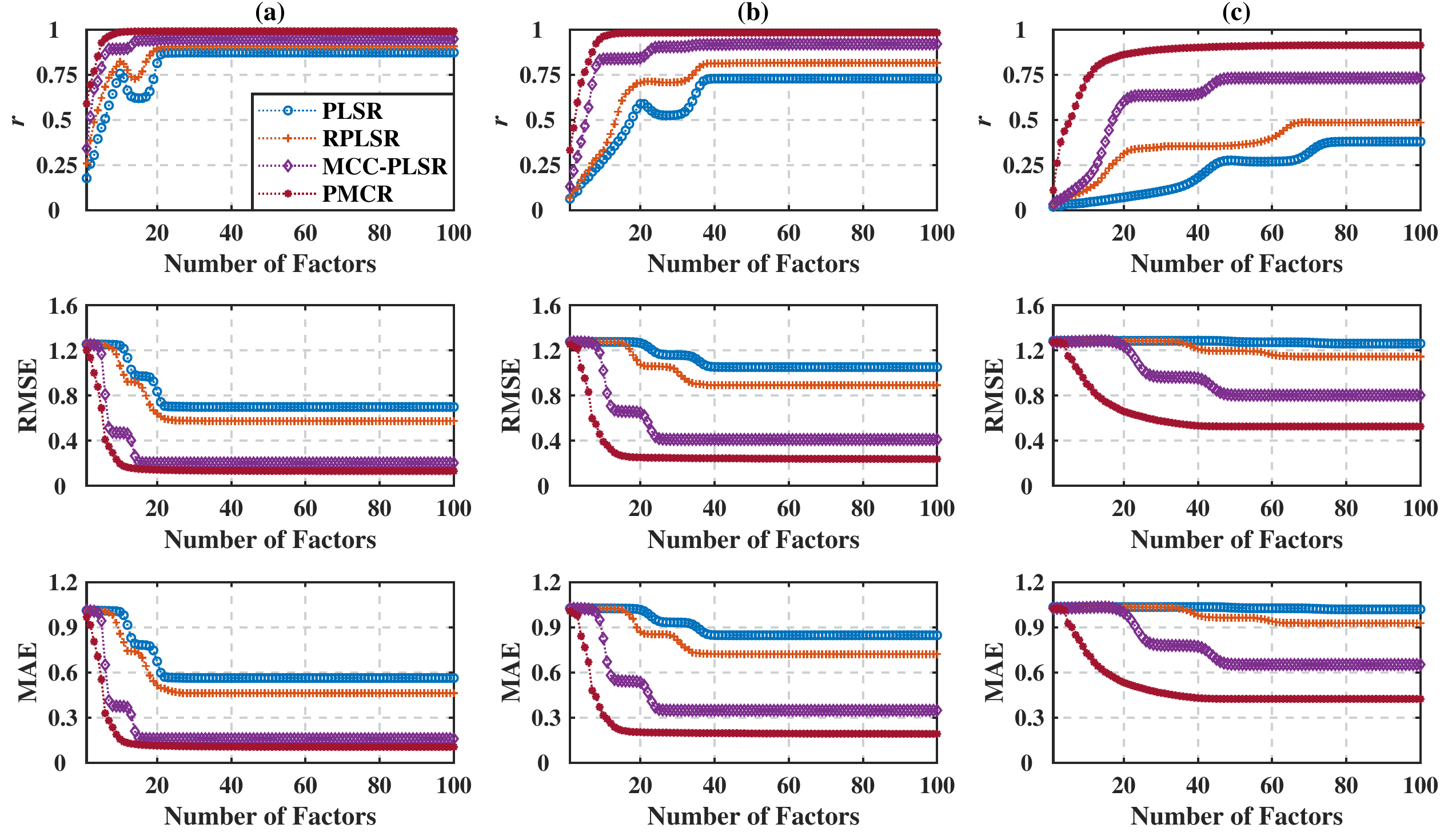}
	\caption{Average regression performance indicators of the synthetic dataset with noise standard deviation being $100$ under three noise levels with the number of factors increasing from $1$ to $100$. (a): noise level $=0.2$, (b): noise level $=0.5$, and (c): noise level $=0.8$. The performance indicators were obtained from $100$ repetitive trials and averaged across three dimensions of the output. The proposed PMCR algorithm not only acquired better prediction results than the other algorithms ultimately with the optimal number of factors, but also achieved admirable regression performance with a small number of factors.}
	\label{fig_n_fac}
\end{figure*}

The number of factors $S$ plays a vital role in PLSR methods, representing the iteration numbers to decompose the input and output matrices. Since it usually causes a notable effect on the result, in addition, we assessed the performance with respect to the number of factors for each method. To this end, we utilized the synthetic dataset from Subsection \ref{subsec:toy} with noise standard deviation being $100$ under three different noise levels, $0.2$, $0.5$, and $0.8$, respectively. The resultant performance indicators for each algorithm with respect to the number of factors are shown in Fig. \ref{fig_n_fac} with $100$ repetitive trials.

One perceives from Fig. \ref{fig_n_fac} that not only the proposed PMCR eventually achieved superior regression performance indicators with the optimal number of factors, but it could realize a rather admirable performance with a small number of factors as well. Specifically, for example, when the noise level was equal to $0.5$ the proposed PMCR achieved its optimal performance with no more than $20$ factors. By comparison, for the other algorithms, when the number of factors was larger than $20$, their regression performance remained promoting significantly. When the noise level was $0.8$, PMCR revealed consistency for all performance indicators when the number of factors became larger than $40$, whereas the other methods acquired their optimal performance with a larger number of factors. This suggests that, PMCR can abstract substantial information with a rather small number of factors from training samples in a noisy regression task.

\subsection{PMCR with Regularization}
\label{subsec:ext_reg_pmcr}
One should additionally note that, the PMCR was proposed by reformulating the conventional PLSR algorithm with using the robust MCC, instead of the mediocre least square criterion. Hence, the proposed PMCR exhibits the supplementary potential for further performance improvements with regularization techniques, as in the existing regularized PLSR methods. The $L_1$-norm regularization could be utilized in (\ref{equ:pmcr_projqrt}) to encourage sparse and robust projectors. In addition, if one requires better smoothness on the output matrix, polynomial or Sobolev-norm penalization could be introduced into PMCR. In the literature, MCC-based algorithm with regularization techniques has been widely investigated. For instance, a robust implementation of the sparse representation classifier (SRC) for face recognition was developed by regularizing the MCC-based SRC loss with $L_1$-norm \cite{he2010maximum}.

\subsection{Extension to Multi-Way Application}
\label{subsec:ext_nway}
The multi-way PLSR establishes the regression relationship between tensor variables with dimensionality reduction by the tensor factorization technique. In the literature, the multi-way PLSR is usually reported to achieve better decoding capability than the generic PLSR algorithm in the brain decoding tasks, where the high-dimensional spatio-spectro-temporal feature is organized with a tensor form. Essentially, the multi-way PLSR decomposes the input and output tensors with the least square criterion by minimizing the Frobenius norm, which represents a generalization of $L_2$-norm \cite{kolda2009tensor}. Hence, the multi-way PLSR is prone to the performance deterioration caused by noises as well.

The proposed PMCR method treats the regression problem of matrix variable, i.e. two-way variable. Extending the PMCR algorithm to multi-way application could probably improve the prediction performance further, which would be investigated in depth in our future works. Promisingly, MCC has been verified effective for tensor variable analysis in a recent study \cite{zhang2016robust}.

\section{Conclusion}
\label{sec:con}
This paper proposed a new robust variant of PLSR algorithm by reformulating the non-robust least square criterion with the sophisticated MCC framework. The proposed robust objective functions can be efficaciously optimized by half-quadratic and fixed-point-based optimization means. Extensive experimental results on the synthetic dataset and Neurotycho epidural ECoG dataset respectively demonstrated that the proposed PMCR can outperform the existing PLSR algorithms, showing promising robustness in high-dimensional and noisy brain decoding tasks.

\ifCLASSOPTIONcaptionsoff
  \newpage
\fi
\bibliography{bibli}
\bibliographystyle{IEEEtran}
\end{document}